\documentclass[aps,prl,floatfix,twocolumn,superscriptaddress,showpacs]{revtex4}
\usepackage{amsmath,epsfig,subfigure,tabularx,multirow}
\newcommand{\lastfootnote}{\footnotemark[\value{footnote}]}

\begin{document}

\title{First-principles approach to rotational-vibrational frequencies and infrared intensity for
H$_2$ adsorbed in nanoporous materials}
\author{Lingzhu Kong}
\affiliation{Dept.\ of Physics \& Astronomy, Rutgers University, Piscataway, NJ 08854, USA}
\author{Yves J. Chabal}
\affiliation{Dept.\ of Materials Science \& Engineering, University of Texas at Dallas,
                     Richardson, TX 75080, USA}
\author{David C. Langreth}
\affiliation{Dept.\ of Physics \& Astronomy, Rutgers University, Piscataway, NJ 08854, USA}
\date{\today}

\begin{abstract}
The absorption sites and the low-lying rotational and vibrational (RV) energy 
states for H$_2$ adsorbed within a metal-organic framework are calculated via 
van der Waals density functional theory. The induced dipole due to bond 
stretching is found to be accurately given by a first-principles driven 
approximation using maximally-localized-Wannier-function analysis. The 
strengths and positions of lines in the complex spectra of RV transitions are 
in reasonable agreement with experiment, and in particular explain the 
experimentally mysteriously missing primary line for para hydrogen.
\end{abstract}

\maketitle
Gas adsorption into nanoporous materials is of great interest for both 
fundamental science and applications. Molecular H$_2$ is challenging because
it can vibrate, rotate, and translate quantum mechanically about its binding 
site due to its small mass. The vibration-rotation (RV) excitations induced 
by infrared (IR) absorption thus provide rich information~\cite{Chabal}.
However, determining the origin and strength of these lines is challenging
because large unit cells are encountered in typical nanoporous structures, 
and the dynamic dipole is distributed over spatially remote parts of the 
structure. To determine the absorption intensity, a precisely tractable 
experimental quantity, one must not only calculate the dipole, but also 
evaluate the quantum mechanical matrix element. An effective approximation 
scheme for doing this has not hitherto been found.

Here, we present such a scheme based on the combination of a self-consistent 
van der Waals density functional (vdW-DF) approach~\cite{vdw} with 
maximally-localized-Wannier-function (MLWF) analysis~\cite{wannier,Marzari} 
and apply it to H$_2$ adsorption in a prototypical metal-organic framework, 
MOF-5~\cite{Yaghi2003}. Such materials have been extensively explored for 
hydrogen storage~\cite{Murray}, gas separation, catalysis,
and sensors~\cite{Czaja}. We analyze the dynamical properties of the 
adsorbed H$_2$, finding results consistent with experiment. Importantly, we
apply the MLWF analysis to calculate the induced dipole moment due to H$_2$ 
adsorption and bond stretching, decomposing the dipole into the contributions 
from both adsorbed dihydrogen and MOF. Monitoring the change in each Wannier 
center of the MOF structure upon H$_2$ adsorption provides an intuitive 
picture by breaking the H$_2$-sorbent interaction into individual components 
of the MOF structure, thus identifying the parts that directly interact with 
the dihydrogen. Such knowledge is important to optimize MOF structures for 
desired properties. In the present case, we use this information to calculate 
the dynamical dipole moment and its matrix element for H$_2$ vibrational 
transitions and RV transitions. We find that the IR intensity of the purely 
vibrational mode for para hydrogen is only about 2.5\% of that for ortho 
hydrogen at the primary adsorption site, which agrees beautifully with the 
missing line in the experiment.~\cite{FitzGerald2008}. A selection rule for RV 
transitions at the relevant site is also obtained and supported by the IR data.

The H$_2$ binding sites are efficiently determined by self-consistent vdW-DF 
calculations~\cite{vdw}. A series of total energy calculations for different 
bond lengths, orientations, and center-of-mass positions respectively are 
performed keeping the MOF atoms fixed at experimental positions~\cite{Rowsell2005}. The resulting potential energy surfaces are then used in the corresponding 
radial and rigid rotor Schr\"{o}dinger equations respectively to extract the 
vibrational, rotational and translational frequencies~\cite{Kongprb,Kongprl}. Anharmonic effects are fully included.

It has been shown that the sum of the Wannier-function centers is connected to 
the Berry phase theory of bulk polarization~\cite{wannier}. The dipole in the 
unit cell is given by
$
\mathbf{u} = e\sum_m Z_m\mathbf{R}_m - e\sum_{n,spin} \mathbf{r}_n
$,
where $Z_m$ and $\mathbf{R}_m$ are the atomic number and position of the 
$m^\mathrm{th}$ nuclei and $\mathbf{r}_n$ is the center of the $n^\mathrm{th}$ 
Wannier function. Importantly, it is trivial to decompose the total dipole 
into components in
various parts of the structure~\cite{Marzari}, which goes beyond the
Berry-phase method. Thereby, we may use the change of Wannier center
upon adsorption as a qualitative measure for understanding the H$_2$-MOF 
interaction and to determine the important parts of the MOF that directly 
interact with hydrogen.

\begin{figure}
\epsfig{file=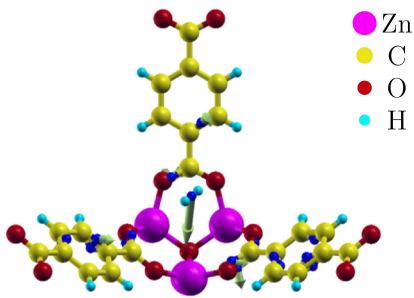,width=2.2in,clip=true}
\caption{ The primary adsorption site and the change of Wannier centers 
(blue balls) due to H$_2$ adsorption compared to bare MOF and free H$_2$.
Vector lengths are enlarged by 1200. The structure shown is just a fraction 
of the unit cell.}
\label{fig:wc_cupsite}
\end{figure}

There are four types of adsorption sites in this structure, as established 
experimentally~\cite{Yildirim2005} and theoretically~\cite{Sillar2009}, 
with reasonable agreement. We start with the positions determined by neutron 
scattering~\cite{Yildirim2005} and relax the H$_2$ with the vdW-DF 
approach, thereby confirming the positions of the four sites, named the 
cup, O3, O2, and benzene sites~\cite{Yildirim2005}. Fig.~\ref{fig:wc_cupsite} 
shows the position of the cup site and a portion of the MOF-5 structure where 
there exists 3-fold rotation symmetry among the three benzene ring branches. 
The distance between the H$_2$ center-of-mass position and the oxygen atom 
passing through the rotation axis is about 4.2\AA\, which is 
somewhat larger than the measured value of 3.8\AA~\cite{Yildirim2005} due to a 
known vdW-DF overestimation of bond lengths~\cite{Langreth2009}. 
Fig.~\ref{fig:wc_cupsite} also shows the shift of the Wannier centers upon 
H$_2$ adsorption with respect to the bare MOF and the free H$_2$. 
See Supplemental Material for other sites \footnote{See Supplemental Material at http://link.aps.org/ supplemental/10.1103/PhysRevLett.000.000000.}.
These figures show that the Wannier centers associated with the 
$\pi$ bonds in the benzene ring change significantly upon H$_2$ adsorption 
for all four adsorption sites, showing a clear and intuitive picture 
of the MOF components that interact directly with the adsorbed H$_2$.

Table~\ref{table:vib} shows both the theoretical and experimental stretch 
frequency shifts of the adsorbed H$_2$ with respect to the corresponding 
free H$_2$ value. The agreement is good, within a reasonable error.
Importantly, the origin of the IR peaks at $-$19 and $-$17 cm$^{-1}$ 
(not understood in the experiment~\cite{FitzGerald2008}) is now unraveled 
with the aid of our calculational results. Note that the calculated binding 
energies at O2 and O3 sites are very close. However, vdW-DF typically 
overestimates intermediate-range interactions~\cite{Lee}. Since O3 has three 
benzene neighbors while O2 has two, the overestimation for the O3 site 
is expected to be larger than that for the O2 site. As a result, the O2 
site is probably more favorable energetically and should get populated more 
than the O3 site. This
is consistent with the measurements where the $-$17 cm$^{-1}$ peak is quite 
weak and appears only as a shoulder to the main line at $-$19 cm$^{-1}$. 
We therefore assign the $-$19 cm$^{-1}$ peak to the O2 site whereas $-$17 
cm$^{-1}$ to O3.

\begin{table}
\caption{ Theory vs experiment~\cite{FitzGerald2008} for the 
stretch frequency shift of the adsorbed H$_2$ relative to free H$_2$. 
See text for zero point energies (not included in the  binding
energies E$_B$ here).}
\label{table:vib}
\begin{tabular*}{0.45\textwidth}{@{\extracolsep{\fill}}cccc}
\hline  \hline

\multirow{2}{*}{site}     &  Theory & Expr. & Calculated E$_B$ \\
                          &  (cm$^{-1}$)       &(cm$^{-1}$)       & (kJ/mol) \\
\hline
 cup     &  $-$23   & $-$27.5  &  $-$11.1 \\
 O2      &  $-$22   & $-$19.0  &  $-$7.9 \\
 O3      &  $-$13   & $-$17    &  $-$7.8 \\
 benzene &  $-$15   &  ---      &  $-$5.4 \\

\hline \hline
\end{tabular*}
\end{table}

\begin{figure} 
{\epsfig{file=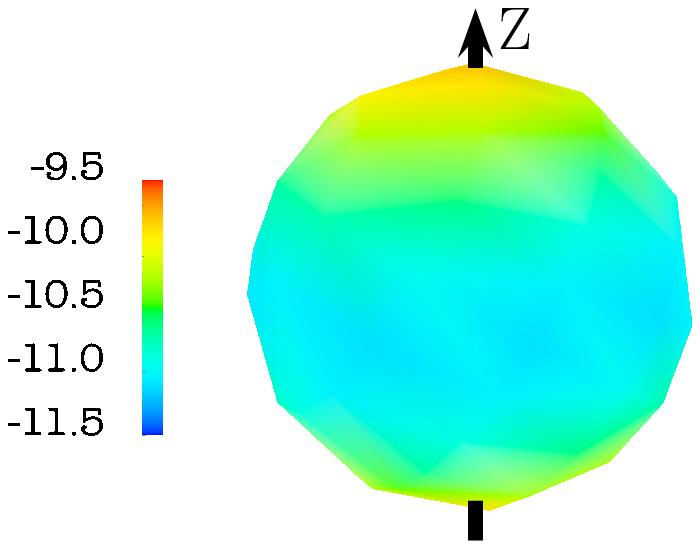,width=.22\textwidth,clip=true} 
 \epsfig{file=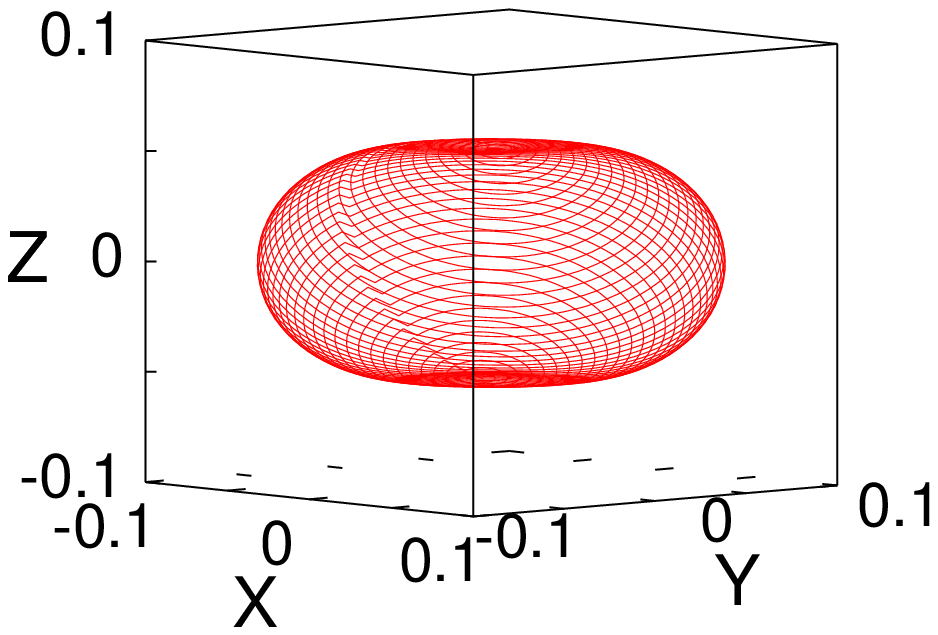,width=.22\textwidth,clip=true} }
\caption{left: Orientational dependence of the binding energy (kJ/mol) at the 
cup site; right: Ground state $|\psi_{rot}|^2$. The distance from the origin
is the probability for that orientation.} 
\label{fig:cup_rotpes}
\end{figure}

The IR spectra also show some 
RV lines where both vibrational and rotational states change during
a single transition. Usually inelastic neutron scattering is employed to study 
the H$_2$ rotational states and has been already applied to H$_2$ in 
MOF-5~\cite{Yaghi2003,Rowselljacs2005}. However, the low energy resolution 
limits detailed analysis. We therefore consider the RV measured with IR
~\cite{FitzGerald2008} for comparison with our rotational calculations.
The left panel of Fig.~\ref{fig:cup_rotpes} is the angular potential 
energy surface at the cup site. The coordinate system is chosen so that the
origin is at the cup site and the $Z$ axis is the 3-fold rotation axis 
(see Fig. S2 in Supplemental Material~\lastfootnote). Fig.~\ref{fig:cup_rotpes} 
shows that H$_2$ tends to lie in the $XY$ plane and to be perpendicular to 
the rotation axis ($Z$). The energies for in-plane orientations are almost 
uniform. Therefore, the rotation is essentially two dimensional, as shown by 
the flattened ground-state angular wave function in the right panel of 
Fig.~\ref{fig:cup_rotpes}. Combining the stretch frequency and the rotational 
energies (see Supplemental Material), we obtain the
RV frequencies. The results for the cup site are shown as S transitions 
in Table~\ref{table:cup_v_intensity}, where the frequency shifts are listed
relative to the corresponding free H$_2$ values
(see Supplemental Material for other sites).
The magnitude of the shifts is consistent between 
theory and experiment, particularly for the leading
peaks in each category that are most intense.

We also calculated the translational frequencies at the cup site associated
with the motion of the whole H$_2$ against the adsorption site. The three 
translational frequencies, at 95, 108 and 133 cm$^{-1}$ respectively, 
are consistent with the value of 84 cm$^{-1}$ extracted from IR 
spectra~\cite{FitzGerald2008}. They are also similar to that observed for 
H$_2$ in C$_{60}$ (110 cm$^{-1}$)~\cite{FitzGerald2002}. The determination
of the rotational and translational states gives the corresponding zero point 
energies of $\sim$0.5 and 2 kJ/mol for H$_2$ at the cup site. 
The binding energy after corrections 
is therefore about 8.5 kJ/mol and somewhat larger than the measured adsorption 
enthalpy of $\sim$5 kJ/mol~\cite{Rowsell2006,Kaye2005}. This 
overestimation by vdW-DF, also found in other MOF 
materials~\cite{Kongprb}, is attributed to overestimation of the 
intermediate-range interactions~\cite{Lee}.

The measured IR spectra for the cup adsorption site shows a strong pure 
vibrational peak due to the ortho-H$_2$, while the corresponding para line 
is not observed. Since the orientational 
energy map only shows a small rotational barrier, the missing para-H$_2$ line 
cannot be explained by the assumption of a frozen H$_2$ orientation. 
Moreover, the local structure around this site has $C_{3v}$ symmetry. 
The rotational state of the para H$_2$ has the same
symmetry as $Z$ and transforms as $A1$. Therefore the transitions between
two $A1$ states should be IR active, even though the X and Y components of 
the dipole give a vanishing contribution by symmetry.

To understand the unexpected missing para-H$_2$ line and 
to calculate the line weights in the more complex RV spectra, we evaluate 
the transition dipole integral explicitly. Assuming the electronic state
remains in the ground state and the RV wave function is separable, one has
$
I_{\alpha} = \langle\psi_{vib}^{f} \psi_{rot}^{f} | u_{\alpha} | \psi_{vib}^{i} \psi_{rot}^{i} \rangle
$,
where $u$ is the dipole moment and $\alpha =X, Y$, or $Z$; the translational 
motion associated with H$_2$ center-of-mass is not included. The dipole is 
a function of the H$_2$ internuclear distance, $R$, and the bond orientation is defined 
by $(\theta,\phi)$. It can be 
expanded as
$
u_{\alpha}(R,\theta,\phi) = u_{\alpha}(R_0,\theta,\phi) + \left. u_{\alpha}'(R,\theta,\phi) \right| _{R_0} \Delta R
$
where $R_0$ is the equilibrium bond length, and $u_{\alpha}'$ is the derivative of 
$u_{\alpha}$ with respect to $R$. Since the vibrational wave functions 
depend only on the inter-nuclear distance, the integral of the first term 
vanishes for transitions between different vibrational states due to 
orthogonality. We find 
$
I_{\alpha} = \langle\psi_{vib}^{f} | \Delta R | \psi_{vib}^{i} \rangle \langle \psi_{rot}^{f} | u_{\alpha}'(R_0,\theta,\phi) | \psi_{rot}^{i}\rangle
\label{eq:I_factor}
$,
where  $| \psi_{rot}^i \rangle\equiv|j_im_i\rangle$ with
$j$ even (odd) for para (ortho) H$_2$, and similarly for $| \psi_{rot}^f \rangle$.
The radial integral is a constant 
for both ortho and para H$_2$ and therefore unnecessary for 
understanding the missing line of para-H$_2$.
The angular integral determines the relative 
intensity between them. We now need to evaluate this integral,
for which $u_{\alpha}'(R_0,\theta,\phi)$ remains to be calculated.

To perform {\sl ab initio} calculations for $u_{\alpha}'$ for every 
($\theta$,$\phi$) is computationally expensive and impractical for this system.
A possible approach is to compute the dipole from first principles for a 
few H$_2$ orientations and derive from them the dipole of all the other 
orientations. This becomes feasible if one can write the dipole as
\begin{equation}
u_{\alpha} = \sum_i C_{i,\alpha} F_{i,\alpha} (\theta,\phi) 
\label{eq:u_ez}
\end{equation}
where $F$ are some known functions and the summation needs to be run over 
only a few terms. 
This approach is appropriate if one realizes that H$_2$ and MOF are weakly 
interacting and the dipole induced on each other can be well described within
a classical picture. First, MOF atoms produce an electric field ($\vec E$)
which induces a dipole on H$_2$. At the cup site, the field is along Z due to the 
rotational symmetry so it can be easily shown that the induced dipole on H$_2$
is of the form in Eq.~\eqref{eq:u_ez} by projecting the field perpendicular and
parallel to the H$_2$ bond and calculating the corresponding dipole components.
A second contribution to the total dipole of the system arises from the H$_2$ 
permanent quadrupole inducing a dipole on the MOF. The quadrupolar
potential and the corresponding electric field at position $\mathbf{r}$,
depend on $\mathbf{r}$, the H$_2$ quadrupole and the bond orientation, which
are again of the form in Eq.~\eqref{eq:u_ez}. This field shifts the MOF 
charge density and induces a dipole. The total dipole on the MOF may be 
formally calculated by multiplying the electric field by the polarizability
at the same position and integrating over the whole MOF. This procedure extends 
the classical picture of point charge into the continuous charge
density regime. It cannot be performed in practice since 
the polarizability is not available. However, the final result for the
dipole would be like the expression in Eq.~\eqref{eq:u_ez},
 since the integration runs over the MOF space while
($\theta,\phi $) would be left unchanged. One can similarly add second-order 
corrections where the induced dipole on H$_2$ and MOF further produce 
dipole on each other. The final equation after this correction turns out
to be quite simple for cup site absorption (see the Supplemental Material for 
derivation) and reads 
\begin{align}
u^s_X &= C_1^s \sin 2 \theta \cos \phi - (C_2^s \cos 2 \phi - C_3^s \sin 2 \phi ) \sin^2 \theta \nonumber \\
u^s_Y &= C_1^s \sin 2 \theta \sin \phi + (C_2^s \sin 2 \phi + C_3^s \cos 2 \phi ) \sin^2 \theta \label{eq:u} \\
u^s_Z &= C_4^s\cos^2 \theta + C_5^s, \nonumber 
\end{align}
where $s$ could be H$_2$, MOF, or the total system. 
The C coefficients depend on the H$_2$ quadrupole, polarizability and 
MOF geometry which are kept fixed during the vibrational 
transition. From Eq.~\eqref{eq:u} we see that only two orientations (each
orientation gives three equations) are required to determine
the five constants and correspondingly the dipole for any other orientations.
To test this model, we calculate $u'_Z$ for several H$_2$ orientations.
Good linearity is obtained between $u'_Z$  and $\cos ^2 \theta$
(see Fig. S7 in Supplemental Material), in agreement with our model. 
X and Y components are also consistent with our model (see 
Supplemental Material).

\begin{table}
\caption{Theory vs.\ experiment~\cite{FitzGerald2008} for 
RV transitions at the cup site.
The  frequency shift $\Delta$v (cm$^{-1}$) is relative to the corresponding free H$_2$
value. The theoretical intensity [$\propto I_\alpha^2$ times the 30K
Boltzmann factor (times 3 for ortho)] is normalized to 100 for the
strongest line; strong (str), weak (wk), and absent (abs) describe the experimental
intensity.}
\label{table:cup_v_intensity}
\begin{tabular*}{0.45\textwidth}{@{\extracolsep{\fill}}lrrrrrr}
\hline  \hline
 & \multirow{2}{*}{$m_i$}
 & \multirow{2}{*}{$m_f$}
 & \multicolumn{2}{c}{Theory}
 & \multicolumn{2}{c}{Experiment} \\
\cline{4-5}
\cline{6-7}
 &  &  &$\Delta$v   &Int.  &$\Delta$v  & Int.  \\
\hline
Q(0)  & 0  & 0
                 & $-23$   & 2 &   & abs  \\
\hline
\multirow{2}{*}{Q(1)} & $\pm1$ & $\pm1$  
&\multirow{2}{*}{$-23$}    & \multirow{2}{*}{97}
&\multirow{2}{*}{$-$27.5}   & \multirow{2}{*}{str}  \\
& 0 & 0 &   &   &  \\
\cline{2-7}
  Q*(1)& $\pm1$ & 0 &22  &9  &39 &wk \\

\hline
 \multirow{3}{*}{S(0)}  &\multirow{3}{*}{0}
   & $\pm$2  &$-$44   & 58 &$-$49.3& str \\
 & & $\pm$1                       &$-12$    & 5 &$-$6.8  & wk  \\
 & & 0                           &$-$1      & 2 &        & abs \\
\hline
 \multirow{8}{*}{S(1)} &\multirow{4}{*}{$\pm1$}
   & $\pm$3  & $-$34 & 100 &$-$36.8 & str  \\
 & & $\pm$2                      & $-$9   & 6 &$-$0.8  & wk \\
 & & $\pm$1                      &    6   & 9 &   21.6 & wk  \\
 & &   0                         &   11   & 3 &        & abs  \\
\cline{2-7}
 &\multirow{4}{*}{0}
   & $\pm$3  &$-$78 &0  &      &abs  \\
 & & $\pm$2                       &$-$53&3&$-$61 &wk   \\
 & & $\pm$1                       &$-$50 &$\sim$0&      &abs   \\
 & &  0                           &$-$33 &$\sim$0&      &abs   \\
\hline \hline
\end{tabular*}
\end{table}

Table~\ref{table:cup_v_intensity} summarizes our results for H$_2$ at the 
cup site.  First we consider pure vibrations where rotational quantum numbers 
do not change (the Q lines in Table~\ref{table:cup_v_intensity}).
The angular integral ($I_A^2$) for Q(0) (para) is much smaller than that 
for Q(1) (ortho), owing partly to the vanishing of the $X$ and $Y$ components 
of the dipole for the Q(0) transition due to symmetry. This symmetry issue 
also applies to the $|jm>=|10\rangle$$\rightarrow$$|10\rangle$ transition 
in Q(1) so that the integral is about 1/3 of that for the other two 
transitions of Q(1). Additionally, Fig.~\ref{fig:cup_rotpes} shows that para 
H$_2$ has a larger probability to be oriented in $XY$ plane, giving a smaller 
$u'_Z$ upon bond stretching, while the $|10\rangle$ state of ortho H$_2$ 
is $p_z$ like and the H$_2$ bond is mainly perpendicular to $XY$ plane. 
As a result, $u'^2_Z$ for the para state is only about one quarter of that 
for the
$|10\rangle$ state (see Supplemental Material for the integral results of 
each component of $\mathbf{u}'$ for the Q transitions). To get the relative 
intensity between Q(0) and Q(1), we need to consider the population ratio 
between para and ortho hydrogen, which we took to be 1:3.  Also, 
the calculated rotational energy of the $|10\rangle$ state is about 5.5 meV 
higher than that for the $m=\pm1$ states (see Supplemental Material). 
As such, its population is about 13\% of 
that of the $m=1$ or $-1$ state at the experimental temperature of 30K within a
Boltzmann distribution approximation. We can therefore estimate that the 
vibrational intensity for para H$_2$ is about 2.5\% of that for ortho 
H$_2$, hence agreeing with the IR measurement, where the
para line was simply not observed~\cite{FitzGerald2008}.

Table~\ref{table:cup_v_intensity} also shows the results for S lines in the
IR spectrum, where $\Delta j=2$.  First a selection rule of $\Delta m=\pm2$ 
is observed with small probabilities for other transitions. This table also 
predicts that there is one single strong line for each S(0) and S(1) at 
the experimental temperature 30K, with shifts of $-$44 and $-$34 cm$^{-1}$ 
respectively, whereas the S(1) line at $-53$ cm$^{-1}$ should be weak due to 
the low population of the $|10\rangle$ state.  More importantly,
the strong line in each category exhibits the largest frequency 
relative to the free H$_2$ value. These results are in
very good agreement with the IR measurements in Ref.~\onlinecite{FitzGerald2008}
where a single strong S(0) line of $-$49.3 cm$^{-1}$ and a strong S(1) line of 
$-$36.8 cm$^{-1}$ are observed for H$_2$ at the cup site.
A weak S(0) line at $-$6.8 cm$^{-1}$ and two S(1) peaks at $-$0.8 and 
21.6 cm$^{-1}$ are also observed with intensities roughly one order of 
magnitude smaller than that of the corresponding strong line, consistent with 
our calculations. Furthermore, Table~\ref{table:cup_v_intensity} shows that 
the calculated intensities of the two strong S lines and the Q(1) lines 
are comparable, which is also observed~\cite{FitzGerald2010}. We also note 
a peak of $\sim${}$-$61 cm$^{-1}$ shift with an intensity similar to that 
of the S(1) at 
$-$0.8 cm$^{-1}$~\cite{FitzGerald2010}. This peak might arise from the 
$|10\rangle$$\rightarrow|$$3,\pm2\rangle$ transitions with theoretical 
intensity close to those of the other two weak S(1) lines, after the 13\% 
population weight is taken into accout(Table~\ref{table:cup_v_intensity}).
Finally, we discuss the special Q*(1) line ($|1,\pm1\rangle\rightarrow|10\rangle$) ) that is experimentally observed~\cite{FitzGerald2008}.
The calculated $I_A^2$ of this transition is approximately equal to that for
Q(0). However, the population between para and ortho hydrogen is $\sim$1:3 
which makes Q*(1) 3$\sim$4 times stronger and observable. The calculated 
shift of 22 cm$^{-1}$ is quite small compared to the experimental value
of 39 cm$^{-1}$. This is likely due to the neglect of 
rotation-translation coupling, which would probably lower the low rotational
state even more and therefore increase the splitting between the $m$=0
and $m$=$\pm$1 states.

In summary, we have proposed a method that provides an intuitive picture of H$_2$ interaction in complex environments. These techniques provide powerful tools for studying gas adsorption in general.

\def\delete{In summary, we applied a first-principles approach~\cite{vdw,Marzari}
to H$_2$ adsorption in a prototypical nanoporous material, confirming
the adsorption sites, and finding RV frequencies consistent with 
experiment~\cite{FitzGerald2008}. We found via a first-principles driven 
approximation that the transition dipole matrix element could be predicted 
for both the H$_2$ as well as remote parts of the structure. We carried out 
the calculations for all the RV transitions from the ground states of para 
and ortho H$_2$ at the primary adsorption site within the observed frequency 
range~\cite{FitzGerald2008}, finding IR intensities in good agreement with 
experiment. We suggested an intuitive picture of the H$_2$ interaction with 
the structure using the change of Wannier centers as a criterion, 
which is useful for seeing at a glimpse which parts of the structure play
important roles in H$_2$ binding. These techniques provide powerful tools for 
studying gas adsorption in general.}

Supported by DOE Grant No.\ DE-FG02-08ER46491.

%\bibliographystyle{prsty}
%\bibliography{draft}

\onecolumngrid
\setcounter{figure}{0}
\setcounter{equation}{0}
\renewcommand{\thefigure}{S\arabic{figure}}
\renewcommand{\thetable}{S\arabic{table}}
\renewcommand{\theequation}{S\arabic{equation}}

\newpage
\begin{center}
\LARGE \textbf {Supplemental Material}
\end{center}

\section*{Calculational methods}
First-principles calculations based on van der Waals density functional theory 
were performed within the plane-wave implementation of the density functional 
theory in the ABINIT package~\cite{ABINIT}, which we have adapted from the 
Siesta~\cite{Siesta} code to incorporate the van der Waals interaction. 
We adopted Troullier-Martins pseudopotentials~\cite{TM} with a 
gradient-corrected functional. An energy cutoff of 50 Ry and Gamma point 
sampling were used for total energy calculations.

\section*{Vibrational frequency}
\begin{figure}[h]
\subfigure{ \epsfig{file=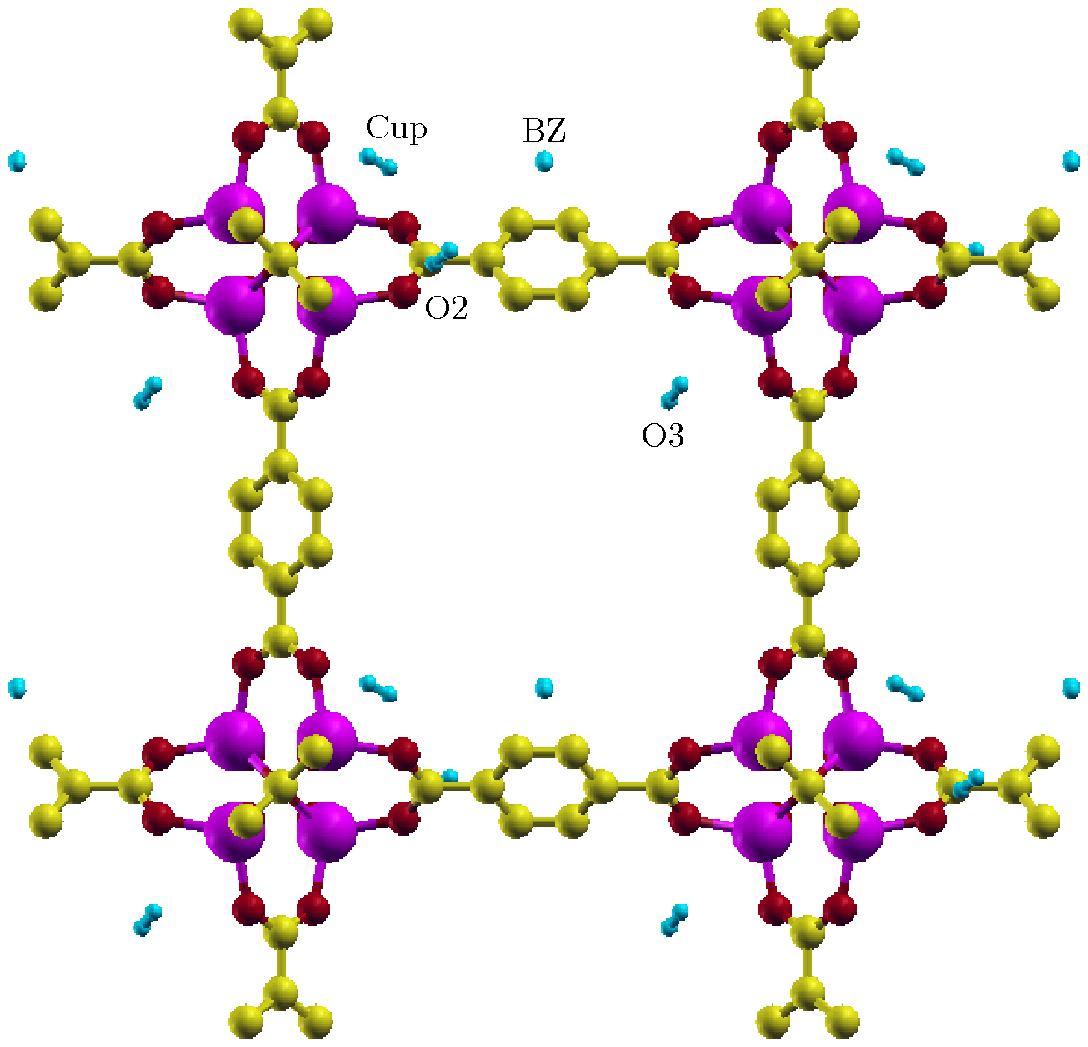,width=3.3in,clip=true}  } \\
\subfigure{ \epsfig{file=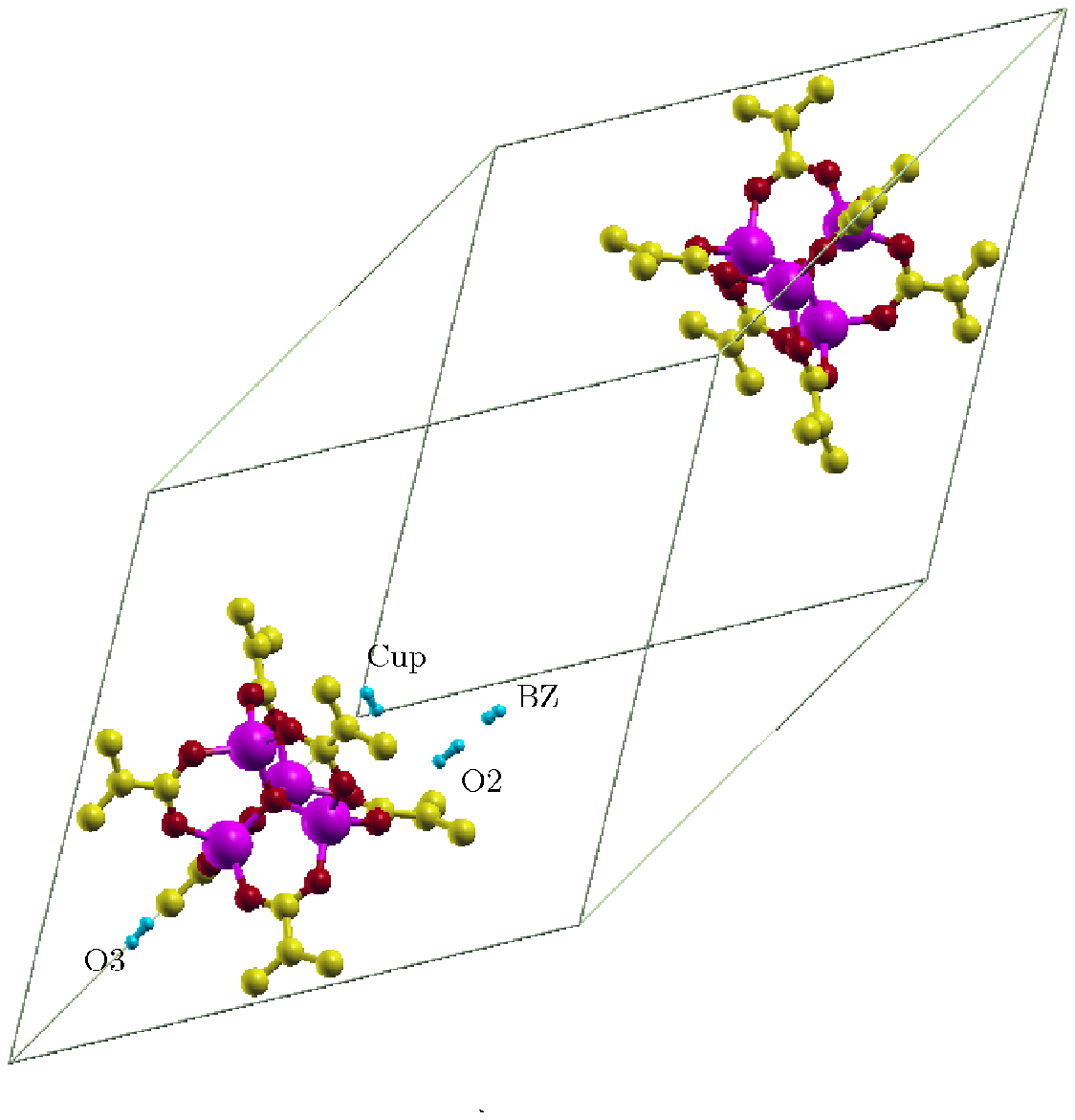,width=3.3in,clip=true}  }
\caption{Illustration of H$_2$ adsorption sites in MOF-5 unit cell (top) and primitive cell (bottom). MOF-5 has FCC structure with lattice constant of 25.89\AA~\cite{Rowselljacs2005}. The primitive cell has 106 atoms. The H atoms on benzene rings are omitted for clarity.}
\label{fig:sites}
\end{figure}

The four type of adsorption sites are shown in Fig.~\ref{fig:sites}.
To calculate the stretch frequency for H$_2$ at each of these four sites, 
we performed a series of calculations 
varying the bond length of H$_2$, with the center of H$_2$ and the host atoms 
fixed at their equilibrium positions. The resulting potential-energy curve 
was used in the Schr{\"o}dinger equation to obtain the eigenvalues and 
excitation frequencies. A similar calculation was also carried out for 
isolated H$_2$ to obtain the frequency shift due to MOF-H$_2$ interaction.
The {\sl ab initio} total energies vs H$_2$ internuclear distance were tabulated
in Tables~\ref{table:cup_vibpes}$-$\ref{table:bz_vibpes} for H$_2$ at the
four type of adsorption sites.

\section*{Rotational frequency}
In order to calculate the rotational states, we first sample the solid angle
to get the total energies. The spherical surface was sampled as follows: the 
polar angle  were evenly divided into seven layers; and 
the azimuthal angle were then sampled by \{1,8,16,24,16,8,1\} number of points
corresponding to each layer from pole to pole of the sphere. We next fit these 
potential energies with spherical harmonics
\begin{equation}
V(\theta,\phi) = \sum_{lm} c_{lm} Y_{lm}
\end{equation}
which was then substituted into the rigid rotor equation and
diagonalized for rotational energies. We found that fitting with $s$ and $d$
states gave results converged within 1 cm$^{-1}$. The fitted coefficients
are shown in Table~\ref{table:ylmcoefficients}. The calculated rotational
energy states are shown in Tables~\ref{table:eigenE_cup}$-$\ref{table:eigenE_bz}.

\section*{Wannier function approach for dipole moment and IR intensity}
\begin{figure}[h]
\epsfig{file=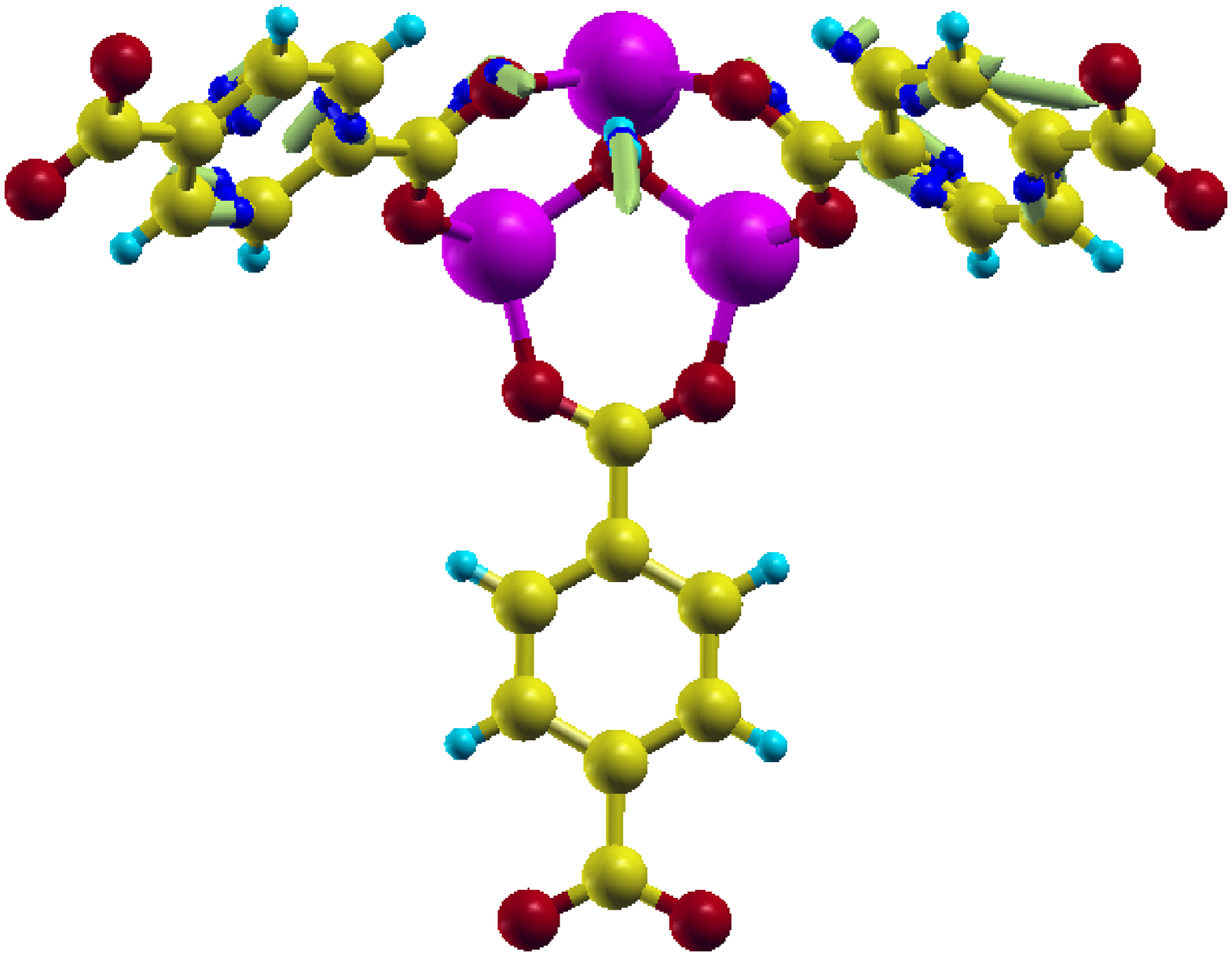,width=4.5in,clip=true}
\caption{Illustration of the O2 adsorption site and the change of Wannier centers due to H$_2$ adsorption compared to bare MOF and free H$_2$. The vector
lengths are enlarged by 1200. }
\label{fig:wc_o2site}
\end{figure}

\begin{figure}[h]
\epsfig{file=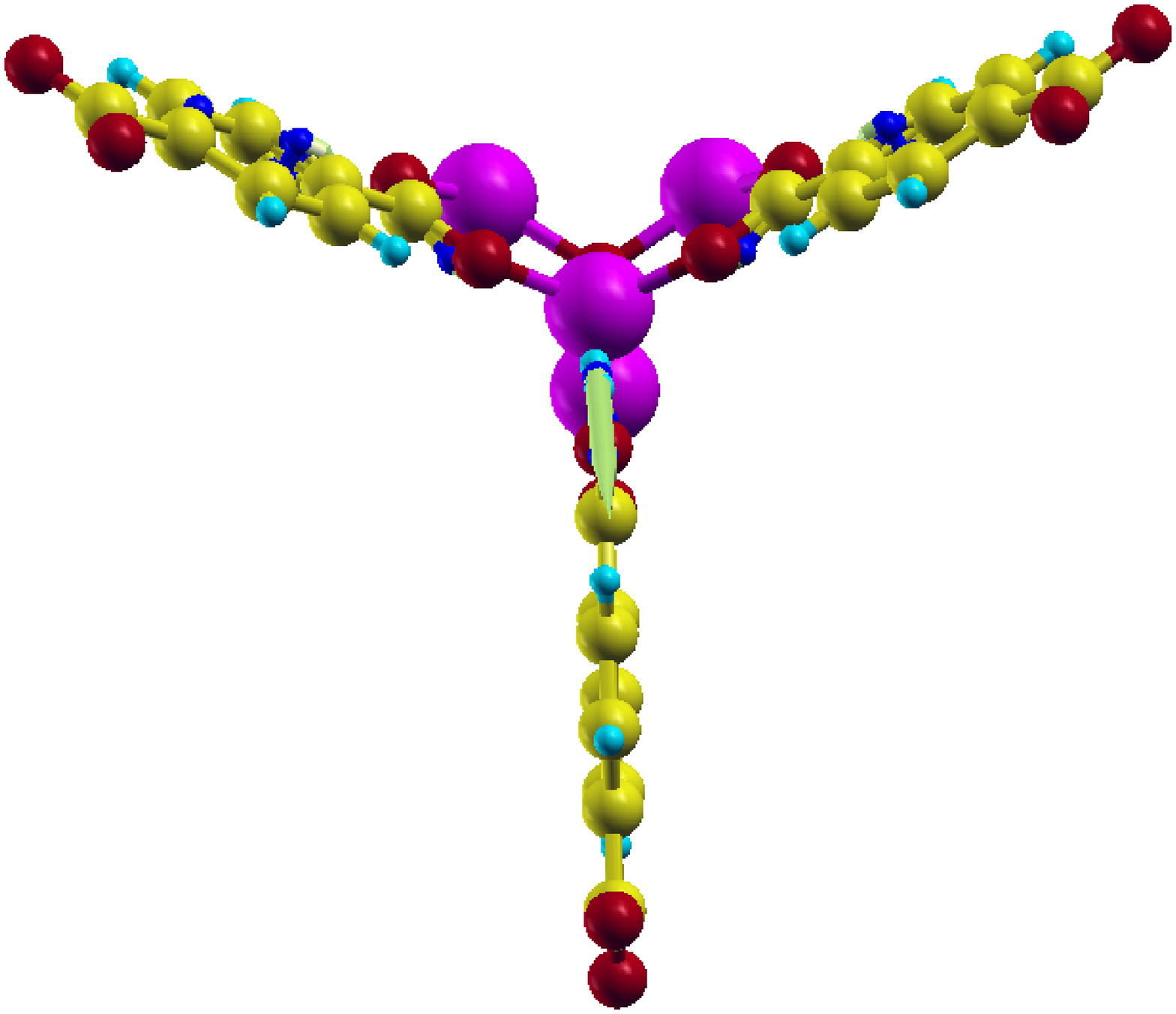,width=4.5in,clip=true}
\caption{Illustration of the O3 adsorption site and the change of Wannier centers due to H$_2$ adsorption compared to bare MOF and free H$_2$. The vector
lengths are enlarged by 1200. }
\label{fig:wc_o3site}
\end{figure}

\begin{figure}[h]
\epsfig{file=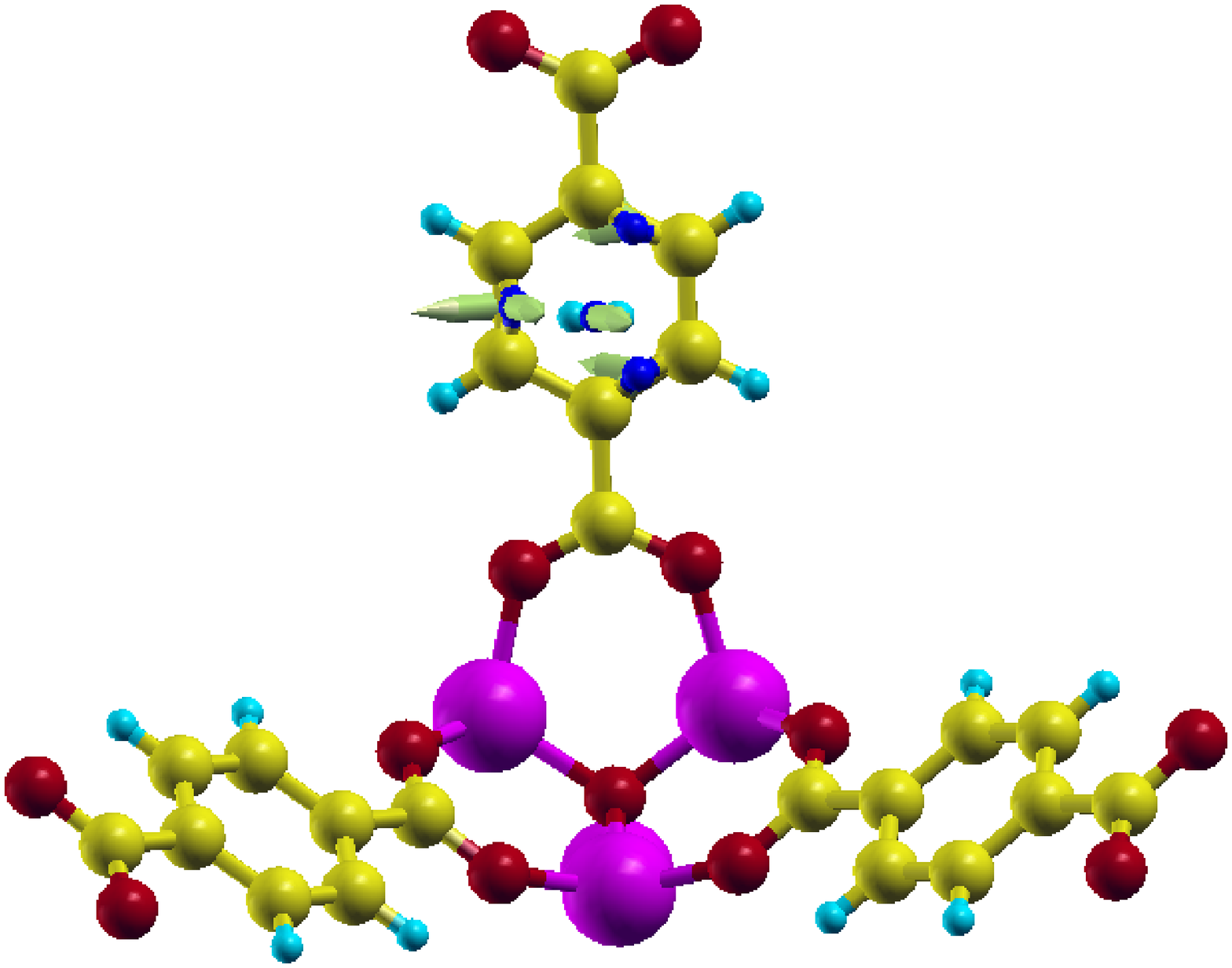,width=4.5in,clip=true}
\caption{Illustration of the benzene adsorption site and the change of Wannier centers due to H$_2$ adsorption compared to bare MOF and free H$_2$. The vector
lengths are enlarged by 1200. }
\label{fig:wc_bzsite}
\end{figure}

The Wannier functions were calculated with the Wannier90 code~\cite{Wannier90}
 embedded in 
ABINIT and the Brillouin zone was sampled by a 2x2x2 Monkhorst-Pack grid. 
The Wannier centers for the bare MOF were first calculated and used as 
initial guess for the H$_2$ loaded system. The change of MOF Wannier centers
upon H$_2$ adsorption was obtained from
\begin{equation}
\delta \mathbf{r}_n =  \mathbf{r}^{MOF+H_2}_n - \mathbf{r}^{MOF}_n
\end{equation}
where $\mathbf{r}_n$ is the center of the $n$-th Wannier function.
Fig.~\ref{fig:wc_o2site}$-$\ref{fig:wc_bzsite} show these changes
for the O2, O3 and benzene sites while the cup site is given in the main text.

\begin{figure}[h]
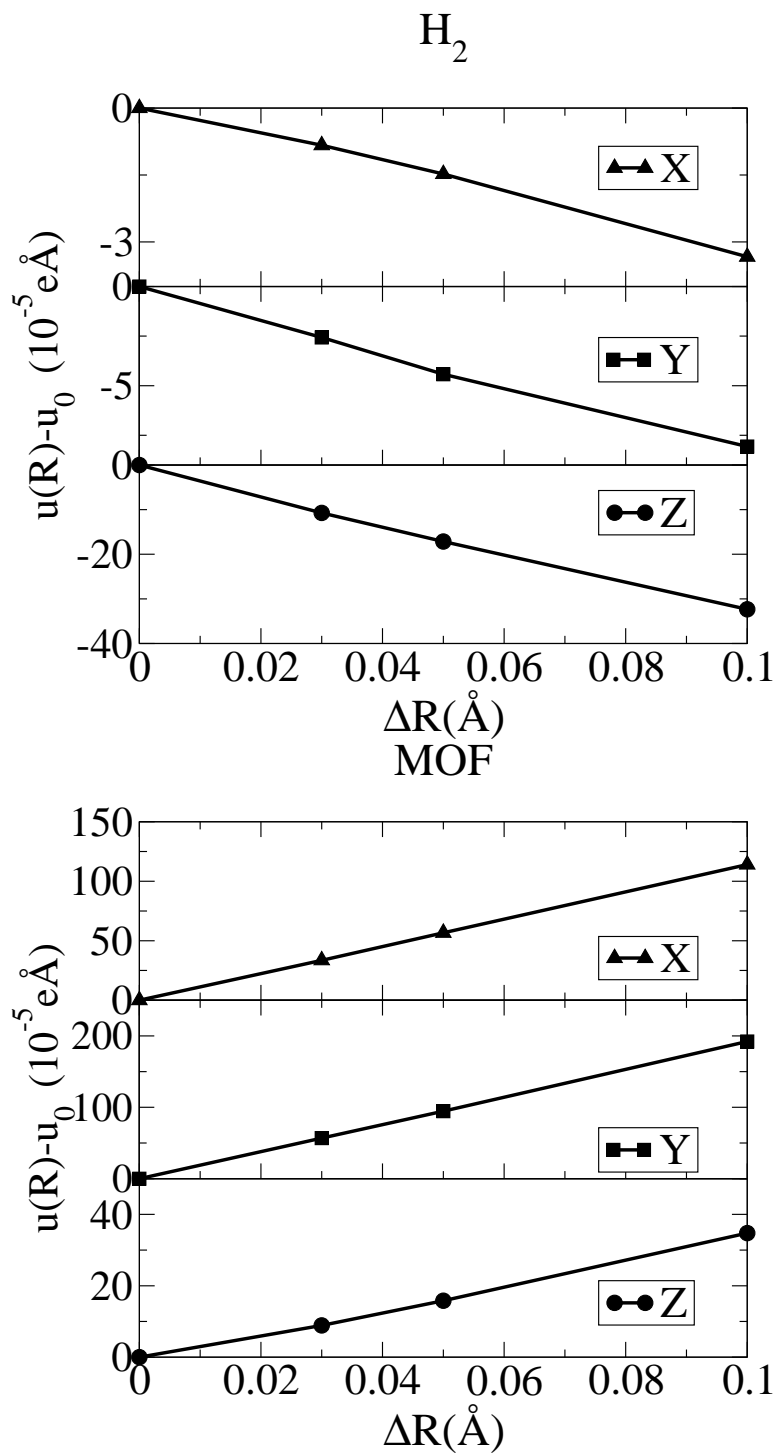

\epsfig{file=./H2_dipole_r.eps,width=4.0in,clip=true}  \\
\epsfig{file=./MOF_dipole_r.eps,width=4.0in,clip=true}
\caption{$\delta {\mathbf u} $ as a function of H$_2$ internuclear distance. u$_0$ is the dipole at equilibrium distance. }\label{fig:der2dif}
\end{figure}

To calculate the IR intensity, one first needs to get the derivative of the 
dipole with respect to the normal coordinates corresponding to H$_2$ stretch 
vibration. 
We used the H$_2$ internuclear distance as an approximation for the stretching 
normal coordinates and the derivative was approximate by the finite difference. 
To reduce the numerical errors, 
the bond stretching should be sufficiently large, but still in the linear 
regime. We found that a stretch of 0.05A from equilibrium bond length was 
appropriate, as shown in Fig.~\ref{fig:der2dif}.

\section*{Model for induced dipole}
The total induced dipole of the system can be approximated by a sum of four 
terms.
\begin{equation}
\mathbf{u} = \mathbf{u}^{H_2}_0 + \mathbf{u}^{MOF}_0 +\mathbf{u}^{MOF}_1 + \mathbf{u}^{H_2}_1
\end{equation}
The first term on the right-hand side is the induced dipole on H$_2$ due 
to interactions with MOF
atoms. The second term is the induced dipole on MOF atoms due to H$_2$
quadrupole. The third term is the induced dipole on MOF due to
$\mathbf{u}^{H_2}_0$ and the fourth term is the induced dipole on H$_2$ due to
$\mathbf{u}^{MOF}_0$. These last two terms are second order corrections. We now 
derive the expressions for the four terms for cup site adsorption.

\subsection{$\mathbf{u}^{H_2}_0$}
Due to the 3-fold symmetry, the electric field ($\vec E$) at cup site due to 
MOF atoms is along the rotation axis, {\sl i.e.} Z. For H$_2$ with its bond 
oriented along ($\theta,\phi$), the projected fields along and perpendicular to
the bond are 
\begin{subequations}
\begin{align}
E_{\|}  & = E\cos \theta \\
E_{\bot}& = E\sin \theta
\end{align}
\end{subequations}
and the corresponding induced dipole is
\begin{subequations}
\begin{align}
u_{\|}  & = E \alpha _{\|} \cos \theta \\
u_{\bot}& = E \alpha _{\bot} \sin \theta
\end{align}
\end{subequations}
where  $\alpha _{\|} $ and  $\alpha _{\bot} $ are the H$_2$ polarizability along and
perpendicular to the bond. In Cartesian coordinates, this gives
\begin{subequations} \label{eq:uH20}
\begin{align}
u_{0X}^{H_2} &= E (\alpha _{\|} - \alpha _{\bot}) \sin \theta \cos \theta \cos \phi  \\
u_{0Y}^{H_2} &= E (\alpha _{\|} - \alpha _{\bot}) \sin \theta \cos \theta \sin \phi  \\
u_{0Z}^{H_2} &= E (\alpha _{\|} \cos ^2 \theta + \alpha _{\bot} \sin ^2 \theta )
\end{align}
\end{subequations}

\subsection{$\mathbf{u}^{MOF}_0$}
Hydrogen molecule has permanent quadrupole. The tensor is 
$Q_{zz}=-2Q_{xx}=-2Q_{yy}=Q$. For H$_2$ at origin with orientation of 
($\theta,\phi $), the quadrupole potential at position P$_1$(X,Y,Z) is
\begin{equation}
V(\mathbf{r}) = \frac{3Q}{2r^5} \tilde{Z} ^2 - \frac{Q}{2r^2}
\end{equation}
where $r=(X^2+Y^2+Z^2)^{1/2}$  and $ \tilde{Z} = X \sin \theta \cos \phi  + Y \sin \theta \sin \phi  + Z \cos \theta  $.   The electric field of this potential is
\begin{subequations}\label{eq:E0X}
\begin{align}
E_{0X}(\mathbf{r}) &= \frac{3Q}{2r^7} \left \{ -2r^2\sin \theta \cos \phi \tilde{Z} + 5X \tilde{Z}^2 \right \} - \frac{3Q}{2r^5}X  \\
E_{0Y}(\mathbf{r}) &= \frac{3Q}{2r^7} \left \{ -2r^2\sin \theta \sin \phi \tilde{Z} + 5Y \tilde{Z}^2 \right \} - \frac{3Q}{2r^5}Y  \\
E_{0X}(\mathbf{r}) &= \frac{3Q}{2r^7} \left \{ -2r^2\sin \theta \cos \phi \tilde{Z} + 5Z \tilde{Z}^2 \right \} - \frac{3Q}{2r^5}Z  
\end{align}
\end{subequations}
The MOF charge density will be shifted by this field and thus leads to induced
dipole.  To calculate this induced dipole on MOF, one may view that there is 
an electron at position P$_1$ with partial charge which is equal to the
charge density at P$_1$. This charge has a certain polarizability. The total 
induced dipole can then be formally calculated by multiplying the electric 
field by the corresponding polarizability and then integrating over the whole 
MOF space. This procedure is somewhat an extension of the classical point charge
into the continuous charge density regime. Assuming the polarizability is 
isotropic, the final result will
have the same dependence on ($\theta,\phi$) as electric field since the 
integration runs over (X,Y,Z) while ($\theta,\phi$) will be left unchanged. In 
other words, we will have an equation of the form in Eq. (5) in the main text.
Note that the isotropic assumption is not critical here except in making the final 
equations simpler. If one had used the whole polarizability tensor, the final
result can still be cast into the form in Eq. (5) in the main text. We found 
that the isotropic assumption gave consistent results for our system.

Due to the 3-fold rotation symmetry, there are three
equivalent points with equal polarizability in MOF. Taking advantage of this
symmetry, the sum of the electric fields at the three positions is
\begin{subequations} \label{eq:EMOF0}
\begin{align}
\tilde E_{0X} & =  \frac{9Q}{4 r^7}\left\{ \left(3 r^2 - 5Z^2 \right) Z \sin 2 \theta \cos \phi           - \frac{5}{2}\left(3XY^2-X^3\right) \sin^2 \theta \cos  2 \phi 
         + \frac{5}{2}\left(3X^2Y-Y^3\right) \sin^2 \theta \sin  2 \phi \right \}  \\
\tilde E_{0Y} & =  \frac{9Q}{4 r^7}\left \{ \left(3 r^2 - 5Z^2 \right)Z \sin 2 \theta \sin \phi  
         + \frac{5}{2}\left(3X^2Y-Y^3\right) \sin^2 \theta \cos  2 \phi 
         + \frac{5}{2}\left(3XY^2-X^3\right) \sin^2 \theta \sin  2 \phi 
                          \right \}  \\
\tilde E_{0Z} & =  \frac{9Q}{4} \frac{ (5Z^2 -3r^2 )Z}{r^7} (3\cos^2 \theta  -1 )
\end{align}
\end{subequations}
The induced dipole on MOF is therefore given by
\begin{subequations} \label{eq:uMOF0}
\begin{align}
u_{0X}^{MOF} & =  C_{01}^{MOF} \sin 2 \theta \cos \phi 
                 -C_{02}^{MOF} \sin^2 \theta \cos  2 \phi
                 +C_{03}^{MOF} \sin^2 \theta \sin  2 \phi   \\
u_{0Y}^{MOF} & =  C_{01}^{MOF} \sin 2 \theta \sin \phi  
                 +C_{03}^{MOF} \sin^2 \theta \cos  2 \phi 
                 +C_{02}^{MOF} \sin^2 \theta \sin  2 \phi   \\
u_{0Z}^{MOF} & = C_{04}^{MOF} (3\cos^2 \theta  -1 )
\end{align}
\end{subequations}
where
\begin{subequations}
\begin{align}
C_{01}^{MOF} &= \int \frac{9Q}{4 r^7} \left(3 r^2 - 5Z^2 \right) Z \alpha^{MOF}(\mathbf{r}) \,d\mathbf{r} \\
C_{02}^{MOF} &= \int \frac{9Q}{4 r^7} \frac{5}{2}\left(3XY^2-X^3\right) \alpha^{MOF}(\mathbf{r}) \,d\mathbf{r} \\
C_{03}^{MOF} &= \int \frac{9Q}{4 r^7} \frac{5}{2}\left(3X^2Y-Y^3\right) \alpha^{MOF}(\mathbf{r}) \,d\mathbf{r} \\
C_{04}^{MOF} &=  \int \frac{9Q}{4r^7} (5Z^2 -3r^2 )Z \alpha^{MOF}(\mathbf{r}) \,d\mathbf{r}
\end{align}
\end{subequations}
and the integration runs over the 1/3 irreducible region of MOF, as a result 
of the 3-fold rotation symmetry.

\subsection{ 2$^{nd}$-order corrections: $\mathbf{u}^{MOF}_1$ and $\mathbf{u}^{H_2}_1$ }
The induced dipole on H$_2$ and MOF, $\mathbf{u}^{H_2}_0$ and $\mathbf{u}^{MOF}_0$, further induces 
dipole on each other and produces second order corrections. For $\mathbf{u}^{H_2}_0$, it gives electric field at P$_1$(X,Y,Z) 
\begin{subequations}
\begin{align}
E_{1X}(\mathbf{r}) &= \frac{1}{r^5}\left(-u^{H_2}_{0X} r^2 +3\mathbf{u}^{H_2}_0 \cdot \mathbf{r} X \right) \\
E_{1Y}(\mathbf{r}) &= \frac{1}{r^5}\left(-u^{H_2}_{0Y} r^2 +3\mathbf{u}^{H_2}_0 \cdot \mathbf{r} Y \right) \\
E_{1Z}(\mathbf{r}) &= \frac{1}{r^5}\left(-u^{H_2}_{0Z} r^2 +3\mathbf{u}^{H_2}_0 \cdot \mathbf{r} Z \right)
\end{align}
\end{subequations}
Similar to the derivation of $\mathbf{u}^{MOF}_0$, one obtains
\begin{subequations}\label{eq:uMOF1}
\begin{align}
u^{MOF}_{1X} &= C_{11}^{MOF} \sin 2 \theta \cos \phi - C_{12}^{MOF} \sin^2 \theta \cos 2 \phi + C_{13}^{MOF} \sin ^2 \theta \sin 2 \phi  \\
u^{MOF}_{1Y} &= C_{11}^{MOF} \sin 2 \theta \sin \phi + C_{12}^{MOF} \sin^2 \theta \sin 2 \phi + C_{13}^{MOF} \sin ^2 \theta \cos 2 \phi  \\
u^{MOF}_{1Z} &= C_{14}^{MOF}\cos^2 \theta + C_{15}^{MOF}
\end{align}
\end{subequations}
where
\begin{subequations}
\begin{align}
C_{11}^{MOF} &= \int \left\{ \frac{9Q}{4}\left(\frac{3Z}{r^5}-\frac{5Z^3}{r^7}\right) +\frac{3E}{4}\left( \frac{1}{r^3}-\frac{3Z^2}{r^5}\right)\left( \alpha_{\|} -\alpha_{\perp} \right)  \right \} \alpha^{MOF} (\mathbf{r}) \,d\mathbf{r} \\
C_{12}^{MOF} &= \int \left\{ \frac{45Q}{8r^7}\left( 3XY^2 -X^3 \right) \right \} \alpha^{MOF} (\mathbf{r}) \,d\mathbf{r}  \\
C_{13}^{MOF} &= \int \left\{ \frac{45Q}{8r^7}\left( 3X^2Y -Y^3 \right) \right \} \alpha^{MOF} (\mathbf{r}) \,d\mathbf{r}  \\
C_{14}^{MOF} &=  \int \left\{ \frac{27Q}{4}\left(-\frac{3Z}{r^5}+\frac{5Z^3}{r^7}\right) - 3E\left(\frac{1}{r^3}-\frac{3Z^2}{r^5}\right)( \alpha_{\|} -\alpha_{\perp} ) \right \} \alpha^{MOF} (\mathbf{r}) \,d\mathbf{r}  \\
C_{15}^{MOF} &= \int \left\{ \frac{9Q}{4}\left(\frac{3Z}{r^5}-\frac{5Z^3}{r^7}\right)  - 3E\left(\frac{1}{r^3}-\frac{3Z^2}{r^5}\right)\alpha_{\perp} \right \} \alpha^{MOF} (\mathbf{r}) \,d\mathbf{r} 
\end{align}
\end{subequations}
and the integration again runs over 1/3 of the MOF region.

Now let us look at the second-order correction on H$_2$, $\mathbf{u}^{H_2}_1$.
The hydrogen quadrupole generates electric field at P$_1$ as given by 
Eq.~\eqref{eq:EMOF0}. With the help of the partial charge and local
polarizability concept, this field produces a local dipole
$ \mathbf{u}^{MOF}_0(\mathbf{r}) =  \alpha^{MOF}(\mathbf{r}) \mathbf{E}_0(\mathbf{r}) $
where $\mathbf{E}_0$($\mathbf{r}$) is given in Eq.~\eqref{eq:E0X} and
 $\alpha^{MOF}(\mathbf{r})$ is assumed to be isotropic.
The electric field back at H$_2$ due to this local dipole is
\begin{subequations}\label{eq:EH2_1}
\begin{align}
E^{H_2}_{1X} & = -\frac{\alpha}{r^5} \left \{ E^{MOF}_{0X} (r^2-3X^2) - 3E^{MOF}_{0Y} X Y - 3E^{MOF}_{0Z} X Z \right \} \\
E^{H_2}_{1Y} & = -\frac{\alpha}{r^5} \left \{ E^{MOF}_{0Y} (r^2-3Y^2) - 3E^{MOF}_{0X} X Y - 3E^{MOF}_{0Z} Y Z \right \} \\
E^{H_2}_{1Z} & = -\frac{\alpha}{r^5} \left \{ E^{MOF}_{0Z} (r^2-3Z^2) - 3E^{MOF}_{0X} X Z - 3E^{MOF}_{0Y} Y Z \right \}
\end{align}
\end{subequations}
Inserting Eq.~\eqref{eq:EMOF0} into Eq.~\eqref{eq:EH2_1}, adding together
the three rotationally equivalent points and integrating over the 1/3 MOF region, one finally obtains
\begin{subequations}\label{eq:EH21}
\begin{align}
\tilde E^{H_2}_{1X} &= C^{H_2}_{11} \sin 2 \theta \cos \phi - C^{H_2}_{12} \sin^2 \theta \cos 2 \phi + C^{H_2}_{13} \sin^2 \theta \sin 2 \phi \\
\tilde E^{H_2}_{1Y} &= C^{H_2}_{11} \sin 2 \theta \cos \phi + C^{H_2}_{12} \sin^2 \theta \sin 2 \phi + C^{H_2}_{13} \sin^2 \theta \cos 2 \phi \\
\tilde E^{H_2}_{1Z} &= C^{H_2}_{14} \cos^2 \theta  + C^{H_2}_{15}
\end{align}
\end{subequations}
where
\begin{subequations}
\begin{align}
C^{H_2}_{11} &= \int  \frac{9\alpha^{MOF}(\mathbf{r}) QZ}{2r^8} \left ( \frac{33}{16} - \frac{Z^2}{8r^2} - \frac{15Z^4}{16r^4} -\frac{5X^2Y^2}{4r^4} \right )  \,d\mathbf{r}\\
C^{H_2}_{12} &= \int   \frac{9\alpha^{MOF}(\mathbf{r}) Q}{2r^{10}}(3Y^2-X^2)X \,d\mathbf{r} \\
C^{H_2}_{13} &= \int  \frac{9\alpha^{MOF}(\mathbf{r}) Q}{2r^{10}}(3X^2-Y^2)Y \,d\mathbf{r}  \\
C^{H_2}_{14} &=  \int \frac{9\alpha^{MOF}(\mathbf{r}) QZ}{2r^8}\left (-\frac{9}{16}+\frac{45}{8}\frac{Z^2}{r^2} +\frac{15}{16}\frac{Z^4}{r^4} + \frac{5X^2Y^2}{4r^4} \right ) \,d\mathbf{r}\\
C^{H_2}_{15} &=  \int \frac{9\alpha^{MOF}(\mathbf{r}) QZ}{2r^8} \left ( \frac{57}{16}-\frac{37}{8}\frac{Z^2}{r^2} -\frac{15}{16}\frac{Z^4}{r^4} -\frac{5X^2Y^2}{4r^4} \right )\,d\mathbf{r}
\end{align}
\end{subequations}
and the integral is over the 1/3 of the MOF space. Considering the anisotropy 
of the polarizability of H$_2$,  we have
\begin{equation} \label{eq:uH21}
 \mathbf{u}^{H_2}_1 \\
 = \left[ \alpha_{\perp}
 + (\alpha_{\|} - \alpha_{\perp})
\left( 
\begin{array}{lll}
  \sin^2 \theta \cos^2 \phi  
& \sin^2 \theta \sin \phi \cos \phi 
& \sin \theta \cos \theta \cos \phi \\
  \sin^2 \theta \sin \phi \cos \phi  
& \sin^2 \theta \sin^2 \phi  
& \sin \theta \cos \theta \sin \phi \\
  \sin \theta \cos \theta \cos \phi  
& \sin \theta \cos \theta \sin \phi 
& \cos^2 \theta
\end{array}
\right) \right]
       \tilde{\mathbf{E}}^{H_2}_1
\end{equation}
The anisotropic term impose a small correction to the first term inside the
bracket. For simplicity, we neglect the second term so that 
$\mathbf{u}^{H_2}_1$ and $\tilde{\mathbf{E}}^{H_2}_1$ have a simple linear 
relationship. In particular, they have the same form of dependence on 
($\theta,\phi$) as given in Eq.~\eqref{eq:EH21}.

\subsection{Coefficients}
\begin{figure}[h]
\epsfig{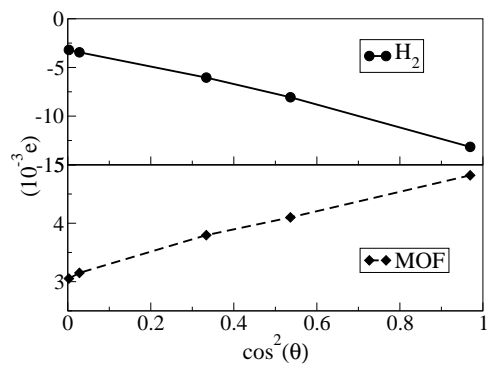}
\caption{$\partial{u_Z}/\partial{R}$ vs $\cos^2 \theta $ for the induced dipole
on H$_2$ and MOF.  $u'_Z$ was calculated as the difference between the dipole
moments at equilibrium bond length and a stretch of 0.05\AA. }
\label{fig:uz_linearity}
\end{figure}

From Eq.~\eqref{eq:uH20}, \eqref{eq:uMOF0}, \eqref{eq:uMOF1} and \eqref{eq:EH21}, we 
conclude that the following equations hold
\begin{subequations}
\begin{align}
u^s_X &= C_1^s \sin 2 \theta \cos \phi - (C_2^s \cos 2 \phi - C_3^s \sin 2 \phi 
) \sin^2 \theta \label{eq:ux} \\
u^s_Y &= C_1^s \sin 2 \theta \sin \phi + (C_2^s \sin 2 \phi + C_3^s \cos 2 \phi 
) \sin^2 \theta \label{eq:uy} \\
u^s_Z &= C_4^s\cos^2 \theta + C_5^s  \label{eq:uz}
\end{align}
\end{subequations}
where $s$ denotes the system and could be H$_2$, MOF or the total. To determine
the coefficients C's, we calculated the dipole and the derivative of the dipole
with respect to H$_2$ internuclear distance with Wannier function approach
for five hydrogen orientations. The Z components of the obtained values were
 used to fit C$_4$ and C$_5$ in Eq.~\eqref{eq:uz}. As shown in 
Fig.~\ref{fig:uz_linearity}, good linearity is obtained in agreement with our
model.

To compute C$_1$, C$_2$ and C$_3$, we pick three {\sl ab initio} calculated value, 
u$'_x$/u$'_y$ of orientation 4 and u$'_x$ of orientation 5, to solve a 
3$\times$3 
linear equation for the coefficients. To check the values obtained,
we substitute them back into Eq.~\eqref{eq:ux} and \eqref{eq:uy} for other 
orientations and compare with the {\sl ab initio} results. The comparison are
shown in Table~\ref{table:compare_deltauxy_MOF} and
\ref{table:compare_deltauxy_H2}. Consistent results are obtained generally
while we do see some deviations on the induced dipole on H$_2$, which may be
due to the neglect of the anisotropy in Eq.~\eqref{eq:uH21}. However, the 
absolute magnitude of these deviations are quite small ($<$ 10\%) compared to 
the total value which is the sum of the induced dipole on MOF and on H$_2$.

\newpage

\newpage
\begin{figure}[h]
\epsfig{file=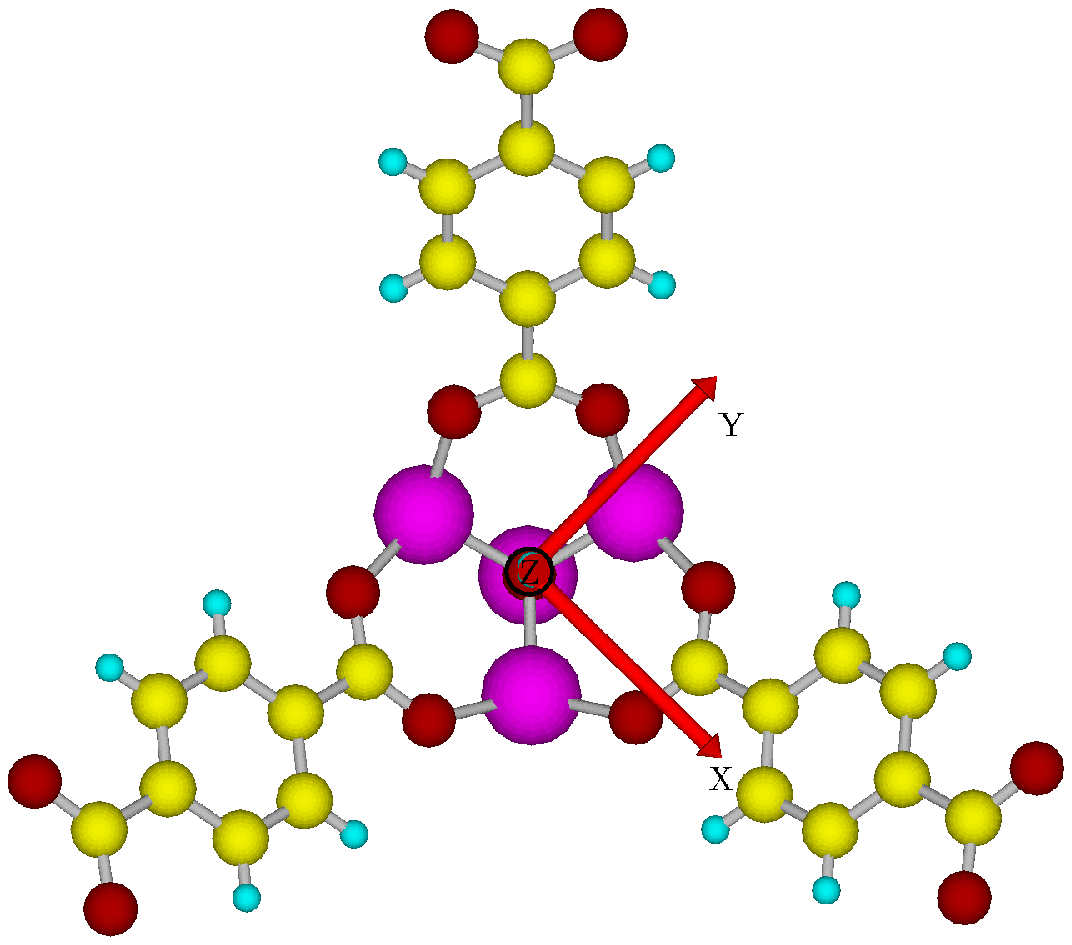,width=4.5in,clip=true}
\caption{Illustration of the MOF frame of reference. Origin is at the cup
adsorption sites and {\sl Z} is along the \textless111\textgreater direction of cubic crystal 
lattice shown in Fig. \ref{fig:sites}. It is also the 3-fold rotation axis.}
\end{figure}

\newpage

\newpage

\newpage

\newpage

\newpage
\begin{table}[h]
\caption{Calculated total energy vs H$_2$ internuclear distance at cup site}
\label{table:cup_vibpes}
\begin{tabular*}{0.5\textwidth}{@{\extracolsep{\fill}}ccc}
\hline \hline
R(a.u.) & E(a.u.)     \\
\hline
0.63258 &  -1147.23459646  \\
0.70817 &  -1147.34182739  \\
0.78376 &  -1147.41767267  \\
0.85935 &  -1147.47155388  \\
0.93494 &  -1147.50972612  \\
1.01052 &  -1147.53646724  \\
1.08611 &  -1147.55478553  \\
1.16170 &  -1147.56683964  \\
1.23729 &  -1147.57419775  \\
1.31288 &  -1147.57801104  \\
1.38847 &  -1147.57913028  \\
1.48297 &  -1147.57768502  \\
1.57745 &  -1147.57390705  \\
1.67194 &  -1147.56843515  \\
1.76642 &  -1147.56176098  \\
1.86091 &  -1147.55423624  \\
1.95540 &  -1147.54614531  \\
2.04988 &  -1147.53771266  \\
2.14437 &  -1147.52910151  \\
2.23886 &  -1147.52044322  \\
2.33334 &  -1147.51183591  \\
\hline \hline
\end{tabular*}
\end{table}

\newpage
\begin{table}[h]
\caption{Calculated total energy vs H$_2$ internuclear distance at O2 site}
\label{table:o2_vibpes}
\begin{tabular*}{0.5\textwidth}{@{\extracolsep{\fill}}ccc}
\hline \hline
R(a.u.) & E(a.u.)     \\
\hline
0.63142 & -1147.23167787 \\
0.70701 & -1147.33946050 \\
0.78259 & -1147.41567703 \\
0.85818 & -1147.46981324 \\
0.93377 & -1147.50816182 \\
1.00936 & -1147.53502878 \\
1.08495 & -1147.55343531 \\
1.16054 & -1147.56555111 \\
1.23613 & -1147.57295236 \\
1.31172 & -1147.57679567 \\
1.38731 & -1147.57793396 \\
1.48182 & -1147.57650225 \\
1.57630 & -1147.57273083 \\
1.67079 & -1147.56726718 \\
1.76528 & -1147.56059280 \\
1.85976 & -1147.55307608 \\
1.95425 & -1147.54499838 \\
2.04873 & -1147.53657940 \\
2.14322 & -1147.52798691 \\
2.23771 & -1147.51934958 \\
2.33219 & -1147.51076491 \\
\hline \hline
\end{tabular*}
\end{table}

\newpage
\begin{table}[h]
\caption{Calculated total energy vs H$_2$ internuclear distance at O3 site}
\label{table:o3_vibpes}
\begin{tabular*}{0.5\textwidth}{@{\extracolsep{\fill}}ccc}
\hline \hline
R(a.u.) & E(a.u.)     \\
\hline
0.63117 & -1147.23131641 \\
0.70676 & -1147.33921650 \\
0.78235 & -1147.41551509 \\
0.85794 & -1147.46970566 \\
0.93353 & -1147.50809162 \\
1.00912 & -1147.53498512 \\
1.08471 & -1147.55340776 \\
1.16030 & -1147.56553250 \\
1.23589 & -1147.57293612 \\
1.31147 & -1147.57677507 \\
1.38705 & -1147.57790325 \\
1.48153 & -1147.57645368 \\
1.57601 & -1147.57265607 \\
1.67050 & -1147.56716130 \\
1.76499 & -1147.56045531 \\
1.85947 & -1147.55290340 \\
1.95396 & -1147.54478993 \\
2.04845 & -1147.53633172 \\
2.14293 & -1147.52769561 \\
2.23742 & -1147.51900844 \\
2.33191 & -1147.51036692 \\
\hline \hline
\end{tabular*}
\end{table}

\newpage
\begin{table}[h]
\caption{Calculated total energy vs H$_2$ internuclear distance at benzene site}
\label{table:bz_vibpes}
\begin{tabular*}{0.5\textwidth}{@{\extracolsep{\fill}}ccc}
\hline \hline
R(a.u.) & E(a.u.)     \\
\hline
0.63097 & -1147.23028851 \\
0.70656 & -1147.33826702 \\
0.78215 & -1147.41460791 \\
0.85774 & -1147.46882070 \\
0.93333 & -1147.50721490 \\
1.00891 & -1147.53410287 \\
1.08450 & -1147.55251525 \\
1.16009 & -1147.56462548 \\
1.23568 & -1147.57201447 \\
1.31127 & -1147.57584010 \\
1.38686 & -1147.57695620 \\
1.48135 & -1147.57549948 \\
1.57583 & -1147.57170006 \\
1.67032 & -1147.56620858 \\
1.76480 & -1147.55950504 \\
1.85929 & -1147.55195560 \\
1.95378 & -1147.54384388 \\
2.04826 & -1147.53538984 \\
2.14275 & -1147.52676285 \\
2.23724 & -1147.51809406 \\
\hline \hline
\end{tabular*}
\end{table}

\newpage
\begin{table}[h]
\caption{ Expansion coefficients (meV) of orientational potential energy 
surface in the basis of spherical harmonics. The equilibrium energy is set to be zero.} 
\label{table:ylmcoefficients}
\begin{tabular*}{0.6\textwidth}{@{\extracolsep{\fill}}ccccccc}
\hline \hline
site & s  &$d_{z^2}$ &  $d_{xz}$ & $d_{yz}$ & $d_{xy}$  & $d_{x^2-y^2}$   \\
\hline
cup     & 17.0  &  0.06 &  8.45 &  8.41  & 8.55  &-0.006  \\
O2      & 25.0  & -3.74 & -6.36 & -6.90  &-5.76  &-4.88   \\
O3      & 7.73  &  0.27 & -2.43 & -2.39  &-2.38  & 0      \\
benzene & 1.46  & -0.52 &  0    &  0.27  & 0     & 0.80   \\
\hline \hline
\end{tabular*}
\end{table}

\newpage
\begin{table}[h]
\caption{Rotational eigen energies (meV) at cup site} 
\label{table:eigenE_cup}
\begin{tabular*}{0.5\textwidth}{@{\extracolsep{\fill}}ccccr}
\hline \hline
State \#  & Energy & E$_i$-E$_1$ & j & m\\
\hline
1 & 4.428 & --       & 0 & 0\\
\hline
2 & 17.498 & 13.070  &\multirow{3}{*}{1}   & $-$1\\
3 & 17.555 & 13.127  &   & 1\\
4 & 23.014 & 18.586  &   & 0\\
\hline
5 & 46.197 & 41.769  &\multirow{5}{*}{2}   & $-$2\\
6 & 46.197 & 41.769  &   & 2\\
7 & 50.094 & 45.666  &   & $-$1\\
8 & 50.135 & 45.707  &   & 1\\
9 & 51.781 & 47.353  &   & 0\\
\hline
10&89.88 & 85.452    & \multirow{7}{*}{3}  & $-$3\\
11&89.88 & 85.452    &   & 3\\
12&92.934 & 88.506   &   & $-$2\\
13&92.934 & 88.506   &   & 2\\
14&94.856 & 90.428   &   & $-$1\\
15&94.894 & 90.466   &   & 1\\
16&95.552 & 91.124   &   & 0\\
\hline \hline
\end{tabular*}
\end{table}

\newpage
\begin{table}[h]
\caption{Rotational eigen energies (meV) at O2 site} 
\label{table:eigenE_o2}
\begin{tabular*}{0.5\textwidth}{@{\extracolsep{\fill}}ccc}
\hline \hline
State \#  & Energy & E$_i$-E$_1$ \\
\hline
1 & 6.76 & -- \\
\hline
2 &18.614 & 11.854 \\
3 &22.306 & 15.546 \\
4 &24.03 & 17.270 \\
\hline
5 &49.036 & 42.276 \\
6 &49.39 & 42.630 \\
7 &50.619 & 43.859 \\
8 &53.265 & 46.505 \\
9 &53.439 & 46.679 \\
\hline
10 &93.026 & 86.266 \\
11 &93.146 & 86.386 \\
12 &94.315 & 87.555 \\
13 &95.211 & 88.451 \\
14 &95.488 & 88.728 \\
15 &97.788 & 91.028 \\
16 &97.801 & 91.041 \\
\hline \hline
\end{tabular*}
\end{table}

\newpage
\begin{table}[h]
\caption{Rotational eigen energies (meV) at O3 site} 
\label{table:eigenE_o3}
\begin{tabular*}{0.5\textwidth}{@{\extracolsep{\fill}}ccccr}
\hline \hline
State \#   & Energy & E$_i$-E$_1$ & j & m\\
\hline
1 &  2.148 & --         & 0 & 0          \\
\hline
2 &    15.815 & 13.667  &\multirow{3}{*}{1} & 0        \\
3 &    17.356 & 15.208  &   & $-$1         \\
4 &    17.434 & 15.286  &   & 1           \\
\hline                            
5 &    45.551 & 43.403  &\multirow{5}{*}{2} &0       \\
6 &    45.869 & 43.721  &   & $-$1        \\
7 &    45.924 & 43.776  &   & 1           \\
8 &    47.026 & 44.878  &   & $-$2        \\
9 &    47.027 & 44.879  &   & 2           \\
\hline                            
10 &    89.685 & 87.537 & \multirow{7}{*}{3} &0       \\
11 &    89.831 & 87.683 &   & $-$1        \\
12 &    89.884 & 87.736 &   & 1           \\
13 &    90.375 & 88.227 &   & $-$2         \\
14 &    90.377 & 88.229 &   & 2        \\
15 &    91.253 & 89.105 &   & $-$3      \\
16 &    91.253 & 89.105 &   & 3         \\
\hline \hline
\end{tabular*}
\end{table}

\newpage
\begin{table}[h]
\caption{Rotational eigen energies (meV) at benzene site} 
\label{table:eigenE_bz}
\begin{tabular*}{0.5\textwidth}{@{\extracolsep{\fill}}ccc}
\hline \hline
State \#   & Energy & E$_i$-E$_1$ \\
\hline
1  & 0.409 & -- \\
\hline
2 & 14.929 & 14.520 \\
3 & 15.051 & 14.642 \\
4 & 15.35 & 14.941 \\
\hline
5 & 44.332 & 43.923 \\
6 & 44.339 & 43.930 \\
7 & 44.553 & 44.144 \\
8 & 44.64 & 44.231 \\
9 & 44.691 & 44.282 \\
\hline
10 &  88.408 & 87.999 \\
11 &  88.409 & 88.000 \\
12 &  88.595 & 88.186 \\
13 &  88.611 & 88.202 \\
14 &  88.693 & 88.284 \\
15 &  88.773 & 88.364 \\
16 &  88.787 & 88.378 \\
\hline \hline
\end{tabular*}
\end{table}

\newpage

\newpage
\begin{table}[t]
\caption{Comparison of ($\delta u_X, \delta u_Y$) on MOF due to H$_2$ bond stretching of 0.05\AA~between the {\sl ab initio} and the fitted values}
\label{table:compare_deltauxy_MOF}
\begin{tabular*}{0.95\textwidth}{@{\extracolsep{\fill}}ccccccc}
\hline \hline
\multirow{2}{*}{orientation} & \multirow{2}{*}{$\theta$(degree)} & \multirow{2}{*}{$\phi$(degree)} & \multicolumn{2}{c}{\sl ab initio}  &  \multicolumn{2}{c}{fitted}   \\
\cline{4-5} \cline{6-7}
                    &&        & $\delta u_x $ & $\delta u_y $  & $\delta u_x $ & $\delta u_y $ \\
\hline
1& 99.650 & -95.723 &  5.690E-04   &   9.440E-04  &   5.690E-04 &  9.634E-04   \\
2& 87.054 & -6.190  & -4.864E-04   &  -8.400E-04  &  -5.178E-04 & -8.720E-04 \\
3& 10.097 & -112.811& -5.760E-05   &  -1.008E-04  &  -5.172E-05 & -9.052E-05 \\
4& 54.741 & -98.278 &  2.774E-04   &   2.464E-04  &   --  &  --  \\
5& 42.937 & -16.295 &  2.640E-04   &  -6.126E-04  &   --  & -5.754E-04  \\
\hline \hline
\end{tabular*}
\end{table}

\begin{table}[h]
\caption{comparison of ($\delta u_X, \delta u_Y$) on H$_2$ due to bond stretching of 0.05\AA~between the {\sl ab initio} and the fitted values}
\label{table:compare_deltauxy_H2}
\begin{tabular*}{0.95\textwidth}{@{\extracolsep{\fill}}ccccccc}
\hline \hline
\multirow{2}{*}{orientation} & \multirow{2}{*}{$\theta$(degree)} & \multirow{2}{*}{$\phi$(degree)} & \multicolumn{2}{c}{\sl ab initio}  &  \multicolumn{2}{c}{fitted}   \\
\cline{4-5} \cline{6-7}
                          && & $\delta u_x $ & $\delta u_y $  & $\delta u_x $ & $\delta u_y $ \\
\hline
1& 99.650 & -95.723  &  -1.631E-05   &  -4.322E-05    &   -2.286E-05  &  -6.552E-05\\
2& 87.054 & -6.190   &   1.562E-05   &   2.220E-05    &    1.340E-05  &   4.522E-05\\
3& 10.097 & -112.811 &   1.338E-05   &   3.052E-05    &    9.392E-06  &   2.030E-05\\
4& 54.741 & -98.278  &  -2.190E-06   &   3.340E-05    & --    &   -- \\
5& 42.937 & -16.295  &  -6.386E-05   &   3.326E-05    & --    &   4.184E-05\\
\hline \hline
\end{tabular*}
\end{table}

\newpage
\begin{table}
\caption{ Theory vs experiment (Ref.~\onlinecite{FitzGerald2008}) for
RV frequency shifts (cm$^{-1}$) of adsorbed H$_2$ relative
to free H$_2$. The vibrational transition is from the ground
state to the 1st excited state ($v$=0 $\rightarrow$ $v$=1). }
\label{table:rovibfrequency}
\begin{tabular*}{0.8\textwidth}{@{\extracolsep{\fill}}ccll}
\hline \hline
\multirow{2}{*}{site} & &S(0) (para) & S(1) (ortho) \\
                     & & ($j$=0$\rightarrow j$=2) & ($j$=1$\rightarrow j$=3) \\
\hline
 \multirow{2}{*}{O2} &Th. &$-37$, $-34$, $-24$, $-3$, $-1$  & $-15$, $-14$, $-4,3$, $5$, $24$, $24$  \\
                     &Ex. &$-36.7$, $-27.3$, $-24.3$, $-7.4$ &$-12.9$ \\

\hline
  O3   &Th. &$-19$, $-16$, $-7$ & $-10$, $-9$, $-4$, $3$   \\
\hline \hline
\end{tabular*}
\end{table}

\begin{table}[h]
\caption{ Angular integral $\langle jm | \mathbf{u'} | jm \rangle $ for H$_2$ at cup site. The energy of the $|00\rangle$ is set as reference. The units for 
$\mathbf{u'}$ is 10$^{-3}$e. }
\label{table:angular_integral}
\begin{tabular*}{0.8\textwidth}{@{\extracolsep{\fill}}cccccc}
\hline  \hline
jm & u$'_X$  & u$'_Y$ & u$'_Z$ & u$'^2(\times 10^{-6}e^2)$ & E(meV) \\
\hline
 00         &   -0.009 &  0.02   & -2.23  & 5.0   & 0 \\
 1$\bar{1}$ &   -2.38  &  7.69   & -1.41  & 66.7  &13.1\\
 11         &    2.36  & -7.67   & -1.40  & 66.4  &13.2\\
 10         &    0.003 &  0.01   & -4.67  & 21.8  &18.6\\

\hline \hline
\end{tabular*}
\end{table}

\newpage
\begin{table}[h]
\caption{Theoretical predictions and experimental data \cite{FitzGerald2008} for
$v=0 \rightarrow v=1$ transitions of H$_2$ at the cup site.
The  frequency shift $\Delta$v (cm$^{-1}$) is relative to the corresponding free
 H$_2$
value and the angular integral given by $I_A^2=|\langle j_fm_f | \mathbf{u'} | j_im_i \rangle|^2$
(10$^{-6}$e$^2$). The rotational energy (meV) $E^{rot}_i$ of the $|00\rangle$ state is set
as a reference. The theoretical intensity is calculated from $I_A^2$ weighted
by the 30K Boltzmann factor and the spin ratio of 1:3 between para
and ortho H$_2$. The strongest line is normalized to 100.} \label{table:cup_v_intensitys}
\begin{tabular*}{0.9\textwidth}{@{\extracolsep{\fill}}ccccccccc}
\hline  \hline & \multirow{2}{*}{m$_i$}
 & \multirow{2}{*}{m$_f$}
 & \multicolumn{4}{c}{Theory}
 & \multicolumn{2}{c}{Experiment} \\
\cline{4-7}
\cline{8-9}
 &  &  &E$_i^{rot}$ &$\Delta$v  & $I^2_A$ &Intensity  &$\Delta$v  & Intensity  \\
\hline
Q(0)($j_i$=0$\rightarrow j_f$=0)  & 0  & 0
                 &0 & -23  & 5 & 2 &   & absent  \\
\hline
\multirow{2}{*}{Q(1)($j_i$=1$\rightarrow j_f$=1)} & $\pm1$ & $\pm1$  &13.1
&\multirow{2}{*}{-23} &66   & \multirow{2}{*}{97}
&\multirow{2}{*}{$-$27.5}   & \multirow{2}{*}{strong}  \\
& 0 & 0 &18.6& &22  &   &  \\
\cline{2-9}
  Q*(1)($j_i$=1$\rightarrow j_f$=1)& $\pm1$ & 0& 13.1 &22 &6 &9  &39 &weak \\

\hline
 \multirow{3}{*}{S(0) ($j_i$=0$\rightarrow j_f$=2)}  &\multirow{3}{*}{0}
   & $\pm$2 &\multirow{3}{*}{13.1} &$-$44 & 115  & 58 &$-$49.3& strong \\
 & & $\pm$1 &                      &$-12$ & 10   & 5 &$-$6.8  & weak  \\
 & & 0      &                      &$-$1  & 5    & 2 &        & absent \\
\hline
 \multirow{8}{*}{S(1)($j_i$=1$\rightarrow j_f$=3)} &\multirow{4}{*}{$\pm1$}
   & $\pm$3 &\multirow{4}{*}{13.1} & $-$34 & 69& 100 &$-$36.8 & strong  \\
 & & $\pm$2 &                      & $-$9  & 4 & 6 &$-$0.8  & weak \\
 & & $\pm$1 &                      &    6  & 6 & 9 &   21.6 & weak  \\
 & &   0    &                      &   11  & 2 & 3 &        & absent  \\
\cline{2-9}
 &\multirow{4}{*}{0}
   & $\pm$3 &\multirow{4}{*}{18.6} &$-$78& 0 &0  &      &absent  \\
 & & $\pm$2 &                      &$-$53& 49&3&$-$61 &weak   \\
 & & $\pm$1 &                      &$-$50& 8 &$\sim$0&      &absent   \\
 & &  0     &                      &$-$33& 6 &$\sim$0&      &absent   \\
\hline \hline
\end{tabular*}
\end{table}

\end{document}